\documentclass{PoS}

\usepackage{graphicx}
\usepackage{dcolumn}
\usepackage{bm}
\usepackage{amsmath,amssymb}

\newcommand{\be}{\begin{equation}}
\newcommand{\ee}{\end{equation}}
\newcommand{\bea}{\begin{eqnarray}}
\newcommand{\eea}{\end{eqnarray}}
\newcommand{\bi}{\begin{itemize}}
\newcommand{\ei}{\end{itemize}}

\newcommand\hyn{\textit{\rm -}}

\title{Charmonium-nucleon interaction from lattice QCD with a relativistic
heavy quark action}

\ShortTitle{Charmonium-nucleon interaction from lattice QCD with a
relativistic heavy quark action}

\author{\textbf{Taichi Kawanai}%
 \thanks{Speaker for Charmonium-nucleon potential from lattice QCD.}\\
Department of Physics, The University of Tokyo, 
Hongo 7-3-1, Tokyo 113-0033, Japan\\
E-mail: \email{kawanai@phys.s.u-tokyo.ac.jp}}

\author{\textbf{Shoichi Sasaki}%
        \thanks{Speaker for Low energy charmonium-nucleon scattering with twisted boundary conditions.}\\
	Department of Physics, The University of Tokyo, 
Hongo 7-3-1, Tokyo 113-0033, Japan\\
       E-mail: \email{ssasaki@phys.s.u-tokyo.ac.jp}}

\abstract{
Detailed information of the low-energy interaction between the charmonia ($\eta_c$ and
$J/\psi$) and the nucleon
is indispensable for exploring the formation of charmonium bound to nuclei.  
In order to investigate the charmonium-nucleon interactions at low energies, 
we adopt two essentially different approaches in lattice QCD simulations.
The charmonium-nucleon potential can be calculated from the equal-time 
Bethe-Salpeter amplitude through the effective Schr\"odinger equation. 
This novel method is based on the same idea originally applied for the nucleon 
force by Aoki-Hatsuda-Ishii.
Another approach is to utilize extended L\"uscher's formula with
partially twisted boundary conditions, which allows us to calculate 
the $s$-wave phase shift at any small value of the relative momentum even in a finite box. 
We then extract model independent information of the scattering length 
and the effective range 
from the phase shift through the effective-range expansion.
Our simulations are carried out 
at a lattice cutoff of $1/a\approx 2$ GeV in a spatial volume of
$(3\;\text{fm})^3$ with the non-perturbatively $O(a)$-improved Wilson fermions 
for the light quarks and a relativistic heavy quark action for the charm quark. 
Although our main results are calculated in quenched lattice calculations, we
also present a preliminary full QCD result by using the 2+1 flavor 
gauge configurations generated by PACS-CS Collaboration. We have found that 
the charmonium-nucleon potential is weakly attractive at short distances 
and exponentially screened at large distances. 
We have also successfully evaluated both the scattering length and 
effective range from the charmonium-nucleon scattering phase shift.
}
\FullConference{The XXVIII International Symposium on Lattice Filed Theory\\
  June 14-19,2010\\
  Villasimius, Sardinia Italy}

\begin{document}

\section{Introduction}
In past several years, properties of hadronic interactions 
have been extensively studied in lattice QCD 
simulations~\cite{Beane:2008ia} based on L\"uscher's finite size method, 
which is proposed as a general method for computing low-energy scattering 
phases of two particles in finite volume~\cite{Luscher:1990ux}.
Here, we recall the recent great success of the nucleon-nucleon potential 
from lattice QCD~\cite{Ishii:2006ec}. In this new attempt, the ``potential'' 
between hadrons can be calculated from the equal-time Bethe-Salpeter (BS) 
amplitude through an effective Schr\"odinger equation~\cite{Aoki:2009ji}.
A direct measurement of {\it hadron-hadron potentials} is now feasible 
by using lattice QCD. We also notice that an idea of partially twisted
boundary conditions, 
which allows us to access any small value of non-zero momentum even in a finite volume,
is quite useful for studying the hadron-hadron interaction at low energies
through L\"uscher's finite size method as originally proposed by Bedaque~\cite{Bedaque:2004kc}. 
In this study, we exploit both novel approaches to obtain detailed information of the low-energy
charmonium-nucleon interaction, that is essential for exploring the formation of charmonium bound to nuclei. 

 Heavy quarkonium states such as charmonium
 ($c\bar{c}$) states do not share the same quark flavor with the nucleon
 ($N$). This suggests that the heavy quarkonium-nucleon interaction is
 mainly induced by the genuine QCD effect of multi-gluon
 exchange~\cite{{Brodsky:1989jd},{Brodsky:1997gh},{Luke:1992tm}}.
 Therefore the $c\bar{c}\hyn N$ system is ideal to study the effect of
 multi-gluon exchange between hadrons. As an analog of the van der
 Waals force, the simple two-gluon exchange contribution gives a weakly
 attractive, but long-ranged interaction
 ~\cite{{Appelquist:1978rt},{Feinberg:1979yw}}. 
 This implies that if such attraction between the charmonium and the
 nucleon is sufficiently strong,
 the charmonia ($\eta_c$ and $J/\psi$) 
 may  be bound to the nucleon or to the large nuclei
 ~\cite{Brodsky:1989jd,Wasson:1991fb}.
 In 1991, Brodsky {\it et al.} had argued that the
 $c\bar{c}$-nucleus ($A$) bound state may be realized for the mass
 number $A\ge 3$, which Wasson confirmed later by solving the
 Schr\"odinger equation for the charmonium-nuclear system with the
 folding potential. 
 Both calculations assumed a simple Yukawa form for the
 charmonium-nucleon potential as $V_{c\bar{c}N}(r)=-\gamma\exp (-\alpha r)/r$
 where parameters ($\alpha=0.6$ GeV, $\gamma=0.6$) are fixed by
 a phenomenological Pomeron exchange model.
 However, the validity of calculations based on a
 phenomenological or perturbative theory is questionable for QCD where
 the strong interaction influences the long distance region.
 
 The $c\bar{c}$-$N$ scattering at low energy has been studied 
 from first principles of QCD. The $s$-wave $J/\psi$-$N$ scattering
 length is about 0.1 fm by using QCD sum
 rules~\cite{Hayashigaki:1998ey} and  $0.71\pm 0.48$ fm ($0.70\pm 0.66$
 fm for $\eta_c$-$N$) by lattice QCD~\cite{Yokokawa:2006td}, while it is
 estimated as large as 0.25 fm from the gluonic van der Waals
 interaction~\cite{Brodsky:1997gh}. All studies suggest that the
 $c\bar{c}$-$N$ interaction is weakly attractive. 
 This indicates that the formation of charmonium bound to nuclei is enhanced.
 In this situation, precise information on the low energy charmonium-nucleon interaction
 is indispensable for exploring nuclear-bound
 charmonium states like the $\eta_c$-${}^{3}{\rm He}$ or
 $J/\psi$-${}^{3}{\rm He}$ bound state in few body
 calculations~\cite{Belyaev:2006vn}.  
 It should be quite important to give a firm theoretical
 prediction about the nuclear-bound charmonium, which is possibly
 investigated by experiments at J-PARC and FAIR/GSI.

\section{Methodology}
Let us briefly review two novel approaches employed in
this study. First, we follow the recent great success of the $N$-$N$
potential from lattice QCD~\cite{Ishii:2006ec}. The potential between
hadrons are calculated from the equal-time BS
amplitude through an effective Schr\"odinger equation~\cite{Ishii:2006ec}.
Second, we exploit partially twisted boundary conditions
to calculate the scattering phase shift
at low energies based on L\"uscher's finite size method.

\subsection{Hadron-hadron potential defined through the BS wave function}

 The method utilized here to calculate the hadron-hadron potential
 in lattice QCD is based on the same idea originally applied for 
 the $N$-$N$ potential~\cite{{Ishii:2006ec},{Aoki:2009ji}}. 
 The first step in the derivation of the hadron-hadron potential
 is to define the BS wave function.
 We calculate the equal-time BS amplitude of two local 
 operators (hadrons $h_1$ and $h_2$) separated by given spatial
 distances ($r=|{\bf r}|$) from the following four-point correlation function
 %
 %
 \begin{equation}
  G^{h_1\hyn h_2}({\bf r}, t; t_2, t_1) =
   \sum_{\bf x}\sum_{{\bf x}^{\prime}, {\bf y}^{\prime}}
   \langle {h_1}({\bf x}, t){h_2}({\bf x}+{\bf r}, t)\left({h_1}
    ({\bf x}^{\prime}, t_2){h_2}({\bf y}^{\prime},t_1)    
 \right)^{\dagger}\rangle,
   \label{Eq.BS_amp}
 \end{equation}
 where ${\bf r}$ is the relative coordinate of two hadrons at sink position ($t$).
 Each hadron operator at source positions ($t_1$ and $t_2$) 
 is separately projected onto a zero-momentum state by a summation 
 over all spatial coordinates ${\bf x}^{\prime}$ and ${\bf y}^{\prime}$.
 To avoid the Fierz rearrangement of two-hadron operators, it is better
 to set $t_2\neq t_1$. 
 Without loss of generality, we choose $t_2=t_1+1=t_{\rm src}$ hereafter.
 Suppose that $|t-t_{\rm src}|\gg 1$ is satisfied, 
 the correlation function asymptotically behaves as 
 \begin{equation}
 G^{h_1\hyn h_2}({\bf r}, t ; t_{\rm src})\propto \phi_{h_1\hyn h_2}({\bf r}) e^{-E_{h_1\hyn h_2}(t-t_{\rm src})} 
 \end{equation}
 where the ${\bf r}$-dependent amplitude $\phi_{h_1\hyn h_2}({\bf r})$,
 which is defined by
\begin{equation}
 \phi_{h_1\hyn h_2}({\bf r})=
 \sum_{{\bf x}}\langle 0| {h_1}
 ({\bf x}){h_2}({\bf x}+{\bf r})|h_1 h_2;E_{h_1\hyn h_2}\rangle,
 \label{Eq.bs_wave}
\end{equation}
with the total energy $E_{h_1 \hyn h_2}$ for the ground state of the two-particle $h_1\hyn h_2$ state,
corresponds to a part of the equal-time BS amplitude and is 
called the BS wave function~\cite{{Luscher:1990ux},{Aoki:2005uf}}.
 After an appropriate projection with respect to discrete rotation
  \begin{equation}
 \phi^{A_{1}^{+}}_{h_1\hyn h_2}(\mathbf{r})=
 \frac{1}{24}\sum_{\mathcal{R}\in O_h} \phi_{h_1\hyn
 h_2}(\mathcal{R}^{-1}\mathbf{r}),
 \label{Eq.projec}
 \end{equation}
 where $\mathcal{R}$ represents 24 elements of the cubic group $O_h$,
 one can get the BS wave function projected in the $A^+_1$ 
 representation, which corresponds to the $s$-wave in continuum theory
 at low energy. 

 The BS wave function defined in Eqs.(\ref{Eq.bs_wave})-(\ref{Eq.projec}) obeys an effective Schr\"odinger
 equation with non-local potential $U_{h_1\hyn h_2}$:
 \begin{equation}
  \left(\frac{1}{2\mu}\nabla^2+E\right)\phi^{A_1^+}_{h_1\hyn h_2}({\bf r})=
   \int U_{h_1\hyn h_2}({\bf r},{\bf r}') \phi^{A_1^+}_{h_1\hyn h_2}({\bf r}')d^3r',
   \label{Eq.non_local}
 \end{equation}
 where {\small$\mu$} and $E$ are a reduced mass and 
 an energy eigenvalue of the two hadron system, respectively.
 The non-local potential $U_{h_1\hyn h_2}$ defined in Eq.(\ref{Eq.non_local}) is supposed to be
 energy independent.
 As long as considering the low energy hadron-hadron scattering, 
 where the relative velocity of hadrons is small, 
 we can take only the leading term in the velocity expansion.
 At low energy, the non-local potential in Eq.(\ref{Eq.non_local}) may become localized as 
 $U_{h_1\hyn h_2}(\mathbf{r},\mathbf{r}')=\delta(\mathbf{r}-\mathbf{r}')V_{h_1\hyn h_2}(\mathbf{r})$.
 As a results, the hadron-hadron ``effective'' central
 potential is defined through the following stationary
 Schr\"odinger equation
 \begin{equation}
  V_{h_1\hyn h_2}(\mathbf{r})=\frac{1}{2\mu}
   \frac{\nabla^2 \phi^{A_1^+}_{h_1\hyn h_2}(\mathbf{r})}
   {\phi^{A_1^+}_{h_1\hyn h_2}(\mathbf{r})}+E.
   \label{Eq.Pot}
 \end{equation}
 Once the BS wave functions $\phi^{A_1^+}_{h_1\hyn h_2}(\mathbf{r})$, 
 the reduced mass $\mu$ and energy eigenvalue $E$
 are calculated in lattice simulations, we can obtain the hadron-hadron
 potential from Eq.(\ref{Eq.Pot}). 
 For the differential operator $\nabla^2$, the discrete Laplacian with nearest-neighbor points is used.
 Although the energy eigenvalue $E$ is supposed to be 
 the energy difference between the total energy of two hadrons
 ($E_{h_1\hyn h_2}$) and the sum of
 the rest mass of individual hadrons ($M_{h_1}+M_{h_2}$), we instead
 determine $E$ with the condition of
 $\lim_{r \rightarrow \infty}\{\frac{1}{2\mu}
   \nabla^2 \phi^{A_1^+}_{h_1\hyn h_2}(\mathbf{r})/
   \phi^{A_1^+}_{h_1\hyn h_2}(\mathbf{r}) + E\}=0$~\cite{Aoki:2005uf}.
 More details of this method can be found in
 Ref.~\cite{Aoki:2009ji}.

\subsection{L\"uscher's finite size method with partially twisted boundary conditions}
\label{Sec:twist}

Let us consider the two-particle system in the center-of-mass frame, where
the total energy of two-particle states is given by 
\begin{equation}
E_{h_1\hyn h_2}(p)=\sqrt{M_{h_1}^2+p^2}+
\sqrt{M_{h_2}^2+p^2}
\label{Eq:IntMom}
\end{equation}
with the relative momentum $p=|{\bf p}|$. We here
introduce the scaled relative momentum, $q=Lp/(2\pi)$,
with the spatial extent $L$. Even under the periodic boundary condition, 
$q^2$ is no longer an integer due to the presence of two-particle interaction.
In this sense, $p$ is the interacting momentum.
The $s$-wave phase shift $\delta_0(p)$ can be calculated
through L\"uscher's phase-shift formula with the interacting momentum measured in the two-particle
system:
\begin{equation}
p\cot \delta_0(p)=\frac{{\cal Z}_{00}(1,q^2)}{L\pi},
\label{Eq:LFSM1}
\end{equation}
where the generalized zeta function, ${\cal Z}_{00}(s, q^2)
=\frac{1}{\sqrt{4\pi}}\sum_{\bf n\in Z^3}({\bf n}^2 - q^2)^{-s}$,
is defined through an analytic continuation in $s$ from the region
$s>3/2$ to $s=1$~\cite{Luscher:1990ux}.
This is a general method for computing low-energy scattering phases of two particles 
in a finite box $L^3$. As is well known, the quantity $p \cot \delta_0(p)$, which appears in the l.h.s. 
of Eq.~(\ref{Eq:LFSM1}), can be expanded in a power series of $p^2$ in the vicinity of the threshold as
\begin{equation}
   p\cot \delta_0(p) = \frac{1}{a_0}+\frac{1}{2}r_0 p^2 +{\cal O}(p^2),
    \label{effecive-range}
  \end{equation}
which is called the effective-range expansion~\cite{Newton:1982qc}. It is worth noting that model-independent information of the low energy interaction should be encoded in a small set of parameters, {\it i.e.} the scattering 
length $a_0$ and the effective range $r_0$, which are associated with the low energy constants
in the effective field theory~\cite{Bedaque:2002mn}.

In principle, one can determine these threshold parameters through the detailed study of 
$p^2$ dependence of the scattering phase shift. However, it should be reminded that accessible 
values of the phase shift on the lattice are restricted due to the discrete momenta (approximately, in units of $2\pi/L$) 
in finite volume. 
Indeed, a typical size of the smallest non-zero momentum under the periodic boundary 
condition, {\it e.g.} $p_{\rm min}\approx 2\pi/L\sim 420$ MeV for $L\simeq 3$ fm and 250 
MeV for $L\simeq 5$ fm, which might be {\it beyond the radius of convergence for 
the effective-range expansion} at least in the attractive interaction case~\footnote{For the $N$-$N$ 
scattering case, the convergence radius of  is known to be smaller than $M_\pi/2$.}.

A novel idea, twisted boundary condition, was proposed by Bedaque 
to circumvent the above mentioned issue~\cite{Bedaque:2004kc}. 
If the following boundary conditions
are imposed on quark fields $q(x)$ in a spatial direction ($i=1,2,3$):
\begin{equation}
q(x_i+L)=e^{i\varphi_i}q(x),
\end{equation}
where $\varphi_i$ represents a twisted angle, 
all momenta in the spatial direction $i$ are quantized in a finite box $L^3$ 
according to
\begin{equation}
p_i=\frac{2\pi}{L}\left(n_i+\frac{\varphi_i}{2\pi}\right),
\end{equation}
where $n_i$ becomes an integer in the free case. 
The case of $\varphi_i=0$ ($\pi$) corresponds to the usual (anti-)periodic boundary condition.
For non-zero twisted angle $\varphi_i\neq 0$, the lowest Fourier mode
($n_i=0$) still can receive non-zero momentum $\varphi_i/L$, which can be set to an arbitrary small value.
The redefinition of quark fields as $q'(x)=e^{i\vec{\theta}\cdot {\vec x}} q(x)$,
where $\vec{\theta}=(\frac{\varphi_1}{L},\frac{\varphi_2}{L},\frac{\varphi_3}{L})$, can suggest 
how to implement the twisted boundary condition.
The new fields $q'(x)$ now satisfy the usual periodic boundary conditions as $q'(x_i+L)=q'(x_i)$. 
Therefore, the hopping terms in the lattice fermion action are transformed \cite{Bedaque:2004kc} as
\begin{equation}
   \sum_{i=1,2,3}\bar{q}'(x)\left[
		e^{ia\theta_i}U_i(x)(1-\gamma_i)q'(x+\hat{i}) 
   + e^{-ia\theta_i}U_i^\dagger(x-\hat{i})(1+\gamma_i)q'(x-\hat{i}) \right].
\end{equation}
This indicates that the quark propagator under the twisted boundary condition 
can be calculated with the simple replacement of the link 
variables $\{U_i(x)\}$ by $\{e^{ia\theta_i}U_i(x)\}$ in the hopping terms~\cite{Bedaque:2004kc,Flynn:2005in}.
The validity of this novel trick has been tested in the dispersion relation of 
single hadron states~\cite{Flynn:2005in,deDivitiis:2004kq}.
It is also widely used for various purposes~\cite{Boyle:2007wg,Boyle:2008yd,Kim:2010sd}.

In this study, we apply twisted boundary conditions to two-hadron system
in order to study properties of two-hadron scattering at low energies
through L\"uscher's finite size method as proposed in the original paper~\cite{Bedaque:2004kc}.
It is should be reminded that the L\"uscher's phase-shift formula receives a slight modification 
under twisted boundary conditions.
The generalized zeta function ${\cal Z}_{00}(s,q^2)$ appeared 
in Eq.~(\ref{Eq:LFSM1}) should be replaced by the following function:
\begin{equation}
   {\cal Z}^d_{00}(s,q^2) = \sum_{\mathbf{n}\in \mathbf{Z}^3}
    \frac{1}{((\mathbf{n}+\mathbf{d})^2-q^2)^s}    
\end{equation}
with $\mathbf{d}=\left(\frac{\varphi_1}{2\pi}, \frac{\varphi_2}{2\pi}, \frac{\varphi_3}{2\pi}\right)$~\cite{Bedaque:2004kc}.
Although large $L$ expansion formula is derived in Ref~\cite{Bedaque:2004kc}
as the asymptotic solution of the new phase-shift formula around the first pole 
of $q^2={\bf d}^2$, we instead use the extended L\"uscher's phase-shift formula
directly. For numerical evaluation of ${\cal Z}^d_{00}(1,q^2)$, we use a
rapid convergent integral expression found in Appendix A of Ref.~\cite{Yamazaki:2004qb}.
Within this approach, it is not necessary to calculate the higher Fourier modes of two-particle scattering state
in order to examine the momentum dependence of the scattering phase shift near the threshold.
It is known that different momentum modes in two-particle states do mix since the relative 
momentum is not conserved due to scattering~\cite{Maiani:1990ca}.
In this sense, there is another advantage of this approach.

\section{Numerical results}

 We have performed lattice QCD simulations in both quenched and full QCD.     
 In quenched QCD, we use a lattice size of $L^3\times T=32^3\times 48$ 
 with the single plaquette gauge action at $\beta=6/g^2=6.0$, which
 corresponds to a lattice cutoff of $a^{-1} \approx 2.1$ GeV according
 to the Sommer scale~\cite{Sommer:1993ce,Guagnelli:1998ud}.
 We use the non-perturbatively ${\cal O}(a)$ improved Wilson fermions 
 for the light quarks ($q$) \cite{Luscher:1996ug} and a relativistic
 heavy quark (RHQ) action for the charm quark ($Q$)~\cite{Aoki:2001ra}.
 The RHQ action is a variant of the Fermilab approach~\cite{ElKhadra:1996mp}, 
 which can remove large discretization errors for heavy quarks. 
 The hopping parameter is chosen to be
 $\kappa_q={0.1342,\ 0.1339,\ 0.1333}$, which correspond to
 $M_\pi={0.64,\ 0.72,\ 0.87}$ GeV ($M_N={1.43,\ 1.52,\ 1.70}$ GeV), 
 and $\kappa_Q=0.1019$ which is
 reserved for the charm-quark mass ($M_{\eta_c}=2.92$
 GeV and $M_{J/\psi}=3.00$ GeV)~\cite{Kayaba:2006cg}. 
 
 Full QCD simulations are also carried out by using 2+1 flavor gauge 
 configurations generated by PACS-CS Collaboration on lattices of size
 $32^3\times 64$
 with the Iwasaki gauge action at $\beta=1.9$, which corresponds to a
 comparable
 lattice cutoff of $a^{-1} \approx 2.2$ GeV, and the non-perturbatively
 ${\cal O}(a)$ 
 improved Wilson fermions with $c_{SW}=1.715$~\cite{Aoki:2008sm}.  
 Although we will present preliminary full QCD results for the $\eta_c
 \hyn N$ potential 
 calculated at the third lightest quark mass
 ($\kappa_{q}=\kappa^{ud}_{\rm sea}=0.13754$) 
 of the PACS-CS configurations~\cite{Aoki:2008sm}, 
 which corresponds to $M_{\pi}=0.41$ GeV
 ($M_N=1.20$ GeV), with $\kappa_Q=0.10679$ for the charm quark
 ($M_{\eta_c}=2.99$ GeV and
 $M_{J/\psi}=3.10$ GeV), our main results are obtained from quenched
 lattice QCD.
 The simulation parameters and the number of sampled
 gauge configurations are summarized in Table~\ref{table1}.
 
We use the conventional interpolating operators,
$h_1(x)=\epsilon_{abc}(u_a(x)C\gamma_5 d_b(x))u_c(x)$ for the nucleon, and
$h_2(y)=\bar{c}_a(y)\gamma_5 c_a(y)$ for the $\eta_c$ state or $h_2(y)=\bar{c}_a(y)\gamma_i c_a(y)$ 
for the $J/\psi$ state, where
$a$,\ $b$ and $c$ are  color
indices, and $C=\gamma_4\gamma_2$ is the charge conjugation matrix. 
Each hadron mass is obtained by fitting the
corresponding two-point correlation function with a single
exponential form. We calculate quark propagators with wall sources,
which are located at $t_{\text{src}}=5$ for the light quarks 
and at $t_{\text{src}}=4$ for the charm quark, with Coulomb gauge
fixing. It is worth mentioning that Dirichlet boundary conditions 
are imposed for quarks in the time direction in order to avoid 
wrap-round effects
which are very cumbersome 
in systems of more than two hadrons~\cite{Takahashi:2005uk}. 
In addition, the ground state dominance in four-point functions is
checked by an effective mass plot of total energies of the $c\bar{c}$-$N$
system.

 \begin{table}[!t]
  \caption{
  Simulation parameters employed in this study. 
  \label{table1}
  }
  \begin{center}
  \begin{tabular}{|cccccccc|}
  \hline
   $n_f$ & $\beta$ & ($\kappa^{ud}_{\rm sea}$, $\kappa_{\rm sea}^{s}$) & $a^{-1}$ [GeV] & $L^3 \times T$ & $La$ [fm] & Stat. & $M_{\pi}$ [GeV] \cr
   \hline
   0 & 6.0 & ---                &2.12 & $32^3\times 48$ & 2.98 & 602 & 0.64-0.87\cr
   2+1 & 1.9 & (0.13754, 0.13640) & 2.18 & $32^3\times 64$ & 2.90 & 450 & 0.41 \cr
   \hline
  \end{tabular}\end{center}
 \end{table}

\subsection{BS wave function and effective central potential}
\begin{figure*}
   \centering
   \includegraphics[width=.49\textwidth]{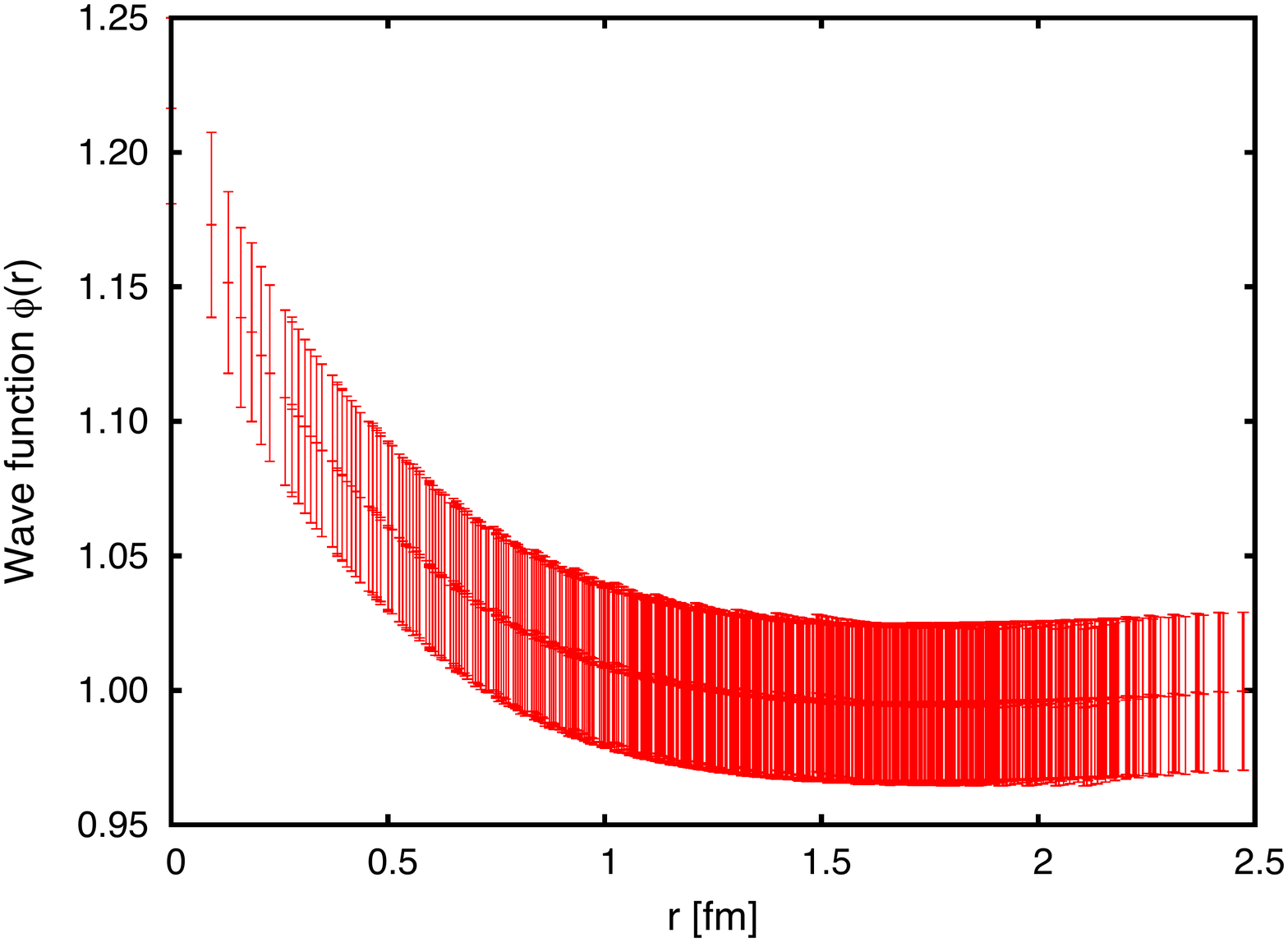}
   \includegraphics[width=.49\textwidth]{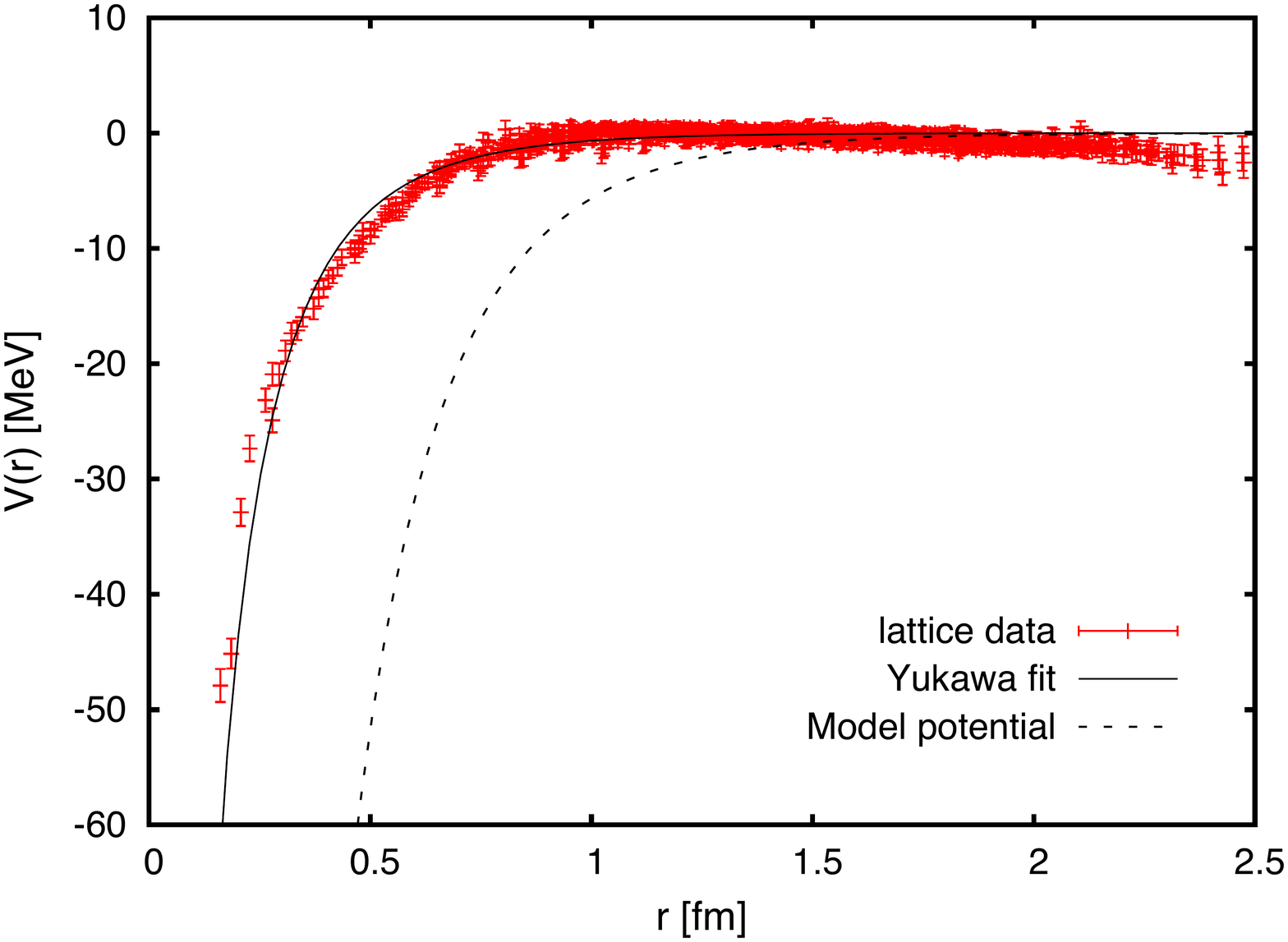}
   \caption{
   The BS wave function (left) and the effective central potential
   (right) in the $s$-wave $\eta_c$-$N$ system for $M_\pi= 0.64$
   GeV as a typical example.  In the right panel we fit a Yukawa
   potential (solid line) and compare with
   the phenomenological potential (dashed) adopted in
   Ref.~\cite{Brodsky:1989jd}.
   \label{Fig_results}}
\end{figure*}

  In this subsection, we mainly show results of the $\eta_c \hyn N$ interaction, 
  which does not possess a spin-dependent part (see Ref.~\cite{Kawanai:2010ev} for 
  results of the $J/\psi \hyn N$ system).
  The left panel of Fig.~\ref{Fig_results} shows a typical result of the
  projected BS wave function at the smallest quark mass in quenched lattice QCD, 
  which is evaluated by a weighted average of data in the time-slice range of $16
  \leq t-t_{\text{src}}\leq 35$. The wave functions are normalized to unity
  at a reference point ${\bf r}=(16,16,16)$, which is supposed to be
  outside of the interaction region. As shown in Fig.~\ref{Fig_results},
  the wave function is enhanced from unity near the origin so that the
  low-energy $\eta_c\hyn N$ interaction is certainly attractive. This
  attractive interaction, however, is not strong enough to form a bound
  state as is evident from this figure, where the wave function is not
  localized, but extends to large distances.
  
  In the right panel of Fig.~\ref{Fig_results}, we show the effective
  central $\eta_c\hyn N$ potential, which is evaluated by the wave
  function through Eq.~(\ref{Eq.Pot}) with measured values of $E$ and $\mu$.
  As is expected, the $\eta_c\hyn N$ potential clearly exhibits an entirely
  attractive interaction between the charmonium and the nucleon 
  without any repulsion at either short or large distances. 
  The short range attraction is deemed to be a result of the absence of
  Pauli blocking,
  that is a relevant feature in this particular system of 
  the heavy quarkonium and the light hadron. The interaction is
  exponentially screened in the long distance region
  $r\gtrsim 1 \text{ fm}$.
  This is consistent with the expected behavior of the color van der Waals
  force in QCD, where
  the strong confining nature of color electric fields must
  emerge~\cite{{Feinberg:1979yw},{Matsuyama:1978hf}}. The
  exponential-type damping in the color van der Waals force is hardly
  introduced by any perturbative arguments. 
  
  In detail, a long-range screening of the color van der Waals force is
  confirmed by the following analysis. We have tried to fit data with two
  types of fitting functions: i) exponential type functions
  $-\exp(-r^m)/r^n$, which include the Yukawa form ($m=1$ and $n=1$), and ii)
  inverse power law  functions $-1/r^n$, where $n$ and $m$ are not
  restricted to be integers. The former case can easily accommodate a
  good fit with a small $\chi^2$/ndf value, while in the latter case we
  cannot get any reasonable fit. For example, the functional forms $-\exp(-
  r)/r$ and  $-1/r^7$ give $\chi^2/\mbox{ndf}\simeq 2.5$ and $34.3$
  , respectively. It is clear that the long range force induced
  by a normal ``van der Waals'' type interaction based on two-gluon
  exchange~\cite{Feinberg:1979yw} is non-perturbatively screened.
  
  If we adopt the Yukawa form $-\gamma e^{-\alpha r}/r$ to fit our data
  of $V_{c\bar{c}\hyn N}(r)$, we obtain $\gamma\sim 0.1$ and $\alpha \sim 0.6$
  GeV. These values should be compared with the phenomenological
  $c\bar{c}$-$N$ potential adopted in Refs.~\cite{{Brodsky:1989jd}},
  where the parameters ($\gamma=0.6$, $\alpha=0.6$ GeV) are barely fixed by a 
  Pomeron exchange model. The strength of the Yukawa potential $\gamma$
  is six times smaller than the phenomenological value, while the Yukawa
  screening parameter $\alpha$ obtained from our data is
  comparable. The $c\bar{c}\hyn N$ potential derived from
  lattice QCD is rather weak.
  
\begin{figure}
  \centering
  \includegraphics[width=.48\textwidth]{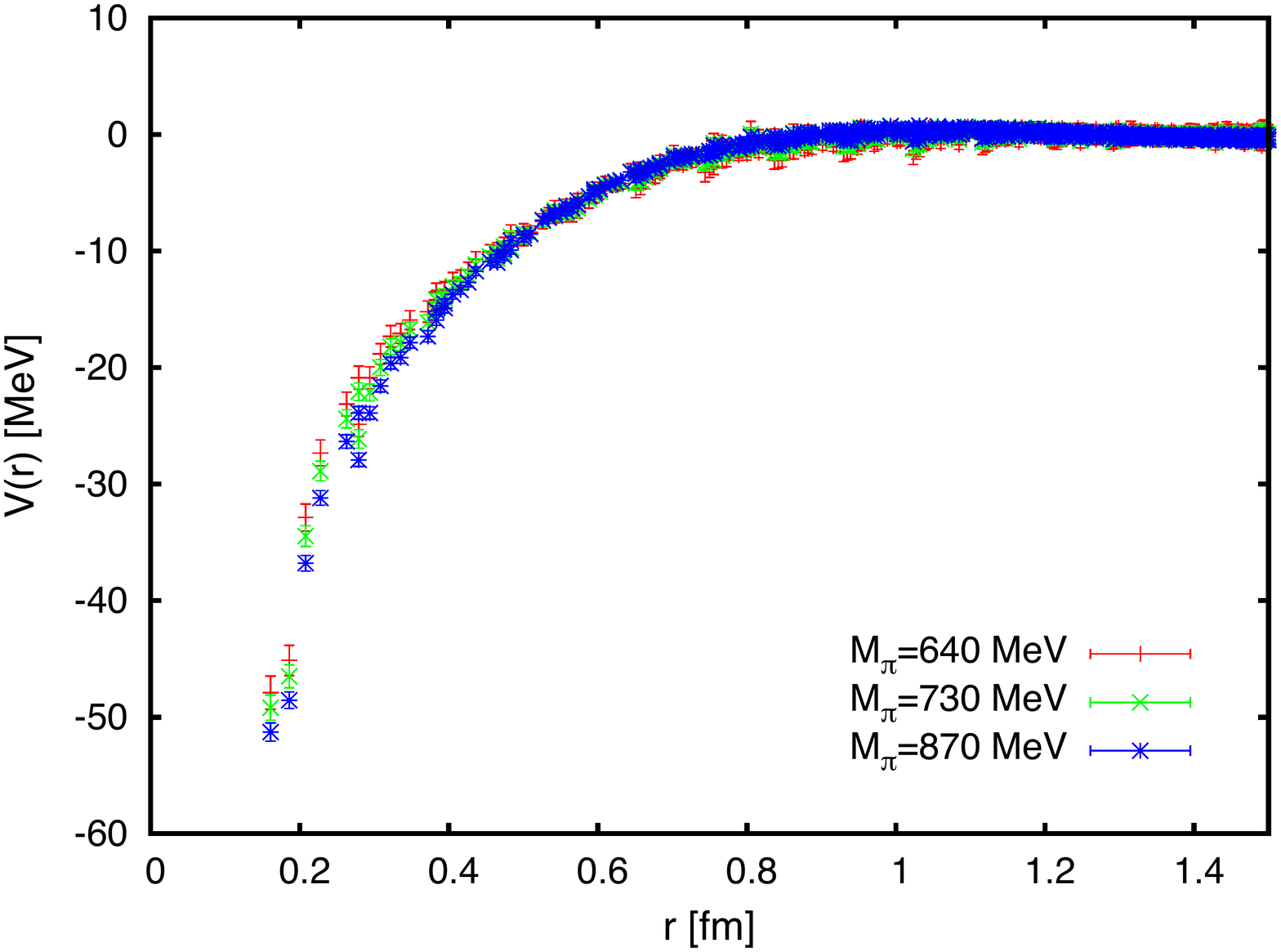}
  \includegraphics[width=.48\textwidth]{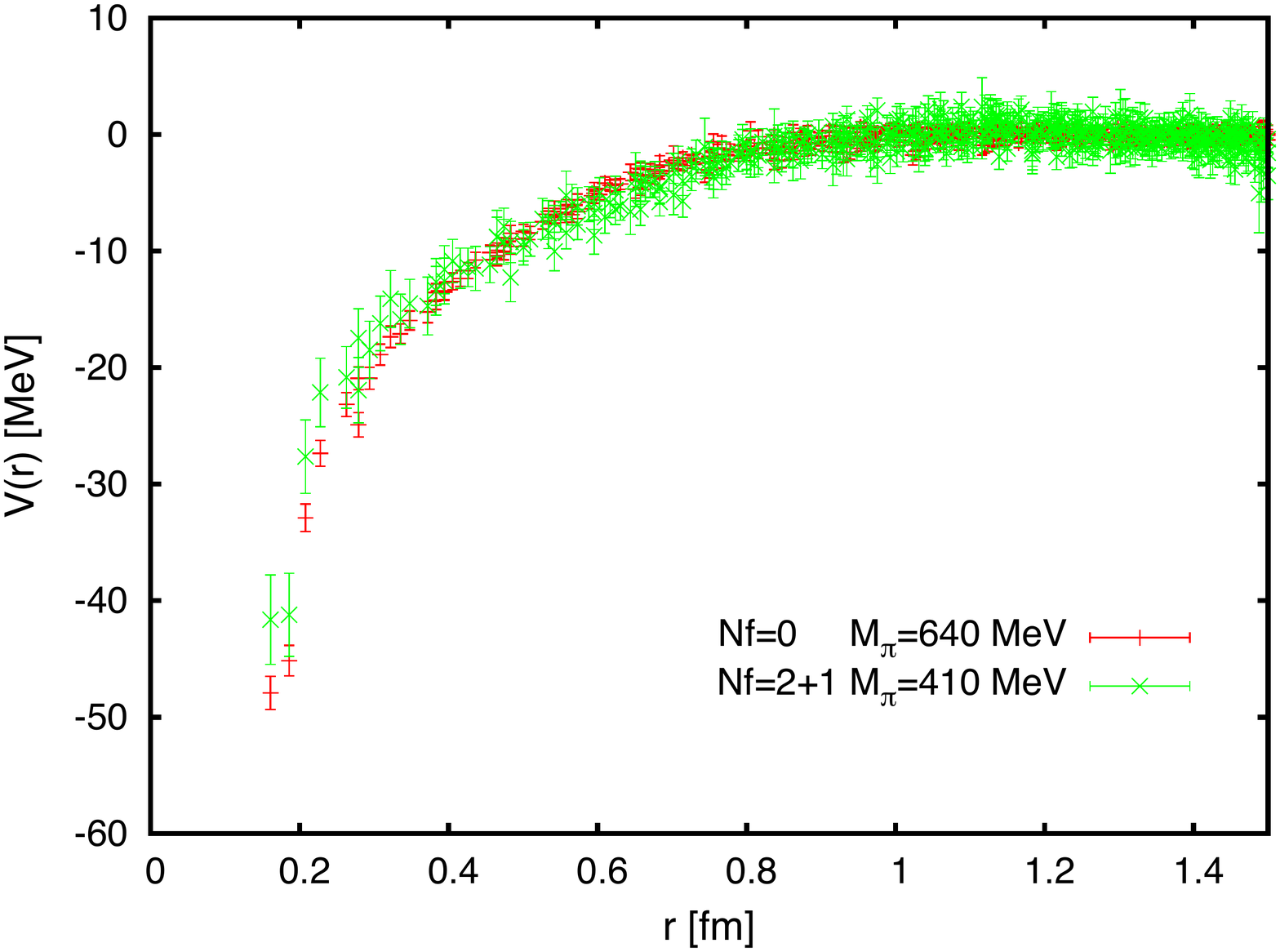}
  \caption{
  The quark-mass dependence of the $\eta_c$-$N$ potential (left) 
  and a comparison between quenched and dynamical simulations (right)
  }
  \label{m_dependence}
  \end{figure}

  We next show the quark-mass dependence of the $\eta_c$-$N$ potential. 
  As shown in the left panel of Fig.~\ref{m_dependence},
  large quark-mass dependence is not observed.
  This is a non-trivial feature since there is an explicit dependence on the
  reduced mass $\mu$ in the definition of the effective central potential
  Eq.(\ref{Eq.Pot}).
  However, if one recalls that the $c\bar{c}\hyn N$ interaction is mainly 
  governed by multi-gluon exchange, the resulting potential is expected to 
  be less sensitive to the reduced mass of the considered system
  ignoring the internal structures of the $\eta_c$ and nucleon states.  

  In order to make a reliable prediction about the nuclear-bound charmonium,
  an important step is clearly an extension to dynamical lattice QCD simulations 
  in the lighter quark mass region. 
  Our preliminary result of the
  $\eta_c$-$N$ potential  at $M_\pi=0.41$ GeV from 2+1 flavor dynamical
  QCD simulation is shown in the right panel of  Fig.~\ref{m_dependence}
  where the $\eta_c$-$N$ potential calculated at $M=0.61$ GeV in quenched
  lattice QCD is also included for comparison.
  There is neither qualitative or quantitative differences between the
  quenched QCD result and the 2+1 flavor QCD result within statistical errors. 
  This indicates that 
  the $\eta_c$-$N$ potential is not strongly affected by dynamical quarks
  at least up to $M_\pi=0.41$ GeV.
  
  It is worth remembering that the ordinary van der Waals interaction is
  sensitive to the size of the charge distribution, which is associated with
  the dipole size. Larger dipole size yields stronger interaction.  
  We may expect that the size of the nucleon becomes large as the light
  quark mass decreases. However, the very mild quark-mass dependence 
  and no appreciable dynamical quark effect
  observed here do not accommodate this 
  expectation properly. 

   Recent detailed studies of nucleon form factors tell us that the root mean-square
   (rms) radius of the nucleon, which is a typical size of the nucleon,
   shows rather mild quark-mass dependence and its value is much smaller than 
   the experimental value up to $M_{\pi}\sim 0.3$ GeV (for example, see~\cite{Yamazaki:2009zq}). 
   At the chiral limit in baryon chiral perturbation theory
   the rms radius is expected to diverge logarithmically~\cite{Beg:1973sc}.
   This implies that the size of the nucleon increases drastically in
   the vicinity of the physical point. It may be phenomenologically regarded 
   as the ``pion-cloud'' effect.
   
  To confirm the possible effect of the nucleon size on 
  the $c\bar{c}$-$N$ potential as described previously,
  we may need to perform simulations in much lighter quark mass region ($M_{\pi} < 0.3$ GeV).
  We speculate that the $c\bar{c}$-$N$ potential would become more attractive
  in the vicinity of the physical point where the ``pion-cloud'' effect emerges.
  Such planning is now underway~\cite{full_QCD}.

\subsection{Scattering under partially twisted boundary conditions}

In this subsection, we present quenched QCD results of low-energy charmonium-nucleon scattering 
with partially twisted boundary conditions. The simulation set up is the same as what we use to calculate
the $\eta_c$-$N$ potential in quenched QCD. In this study, we introduce twisted boundary conditions
only in a single direction, $z$-direction, where the $D_4$ point group symmetry
still remains as a remnant of the rotation symmetry. Three non-zero twisted angles 
are chosen to be $\varphi_3=\alpha, 2\alpha$ and $3\alpha$ with $\alpha = 0.03 \times L \approx 3\pi/10$.

We first examine the dispersion relation of the charmonia ($\eta_c$ and $J/\psi$) 
and the nucleon in order to demonstrate that finite momenta can be properly induced 
for the lowest Fourier mode ($|{\bf n}|=0$) by the twisted angles. For comparing with
results obtained from the higher Fourier modes ($|{\bf n}| \neq 0$) in the periodic boundary condition, 
we additionally calculate quark propagators with the gauge-invariant Gaussian smearing 
source~\cite{Gusken:1989qx}.  We then compute two-point functions of the hadrons, where 
the sink hadron operators are projected onto the three lowest momenta, 
$(0,0,0)$, $(1,0,0)$ and $(1,1,0)$ in units of $2\pi/L$ in the periodic boundary condition
(see details in Ref.~\cite{Sasaki:2007gw}).

\begin{figure*}
  \centering
  \includegraphics[width=.48\textwidth]{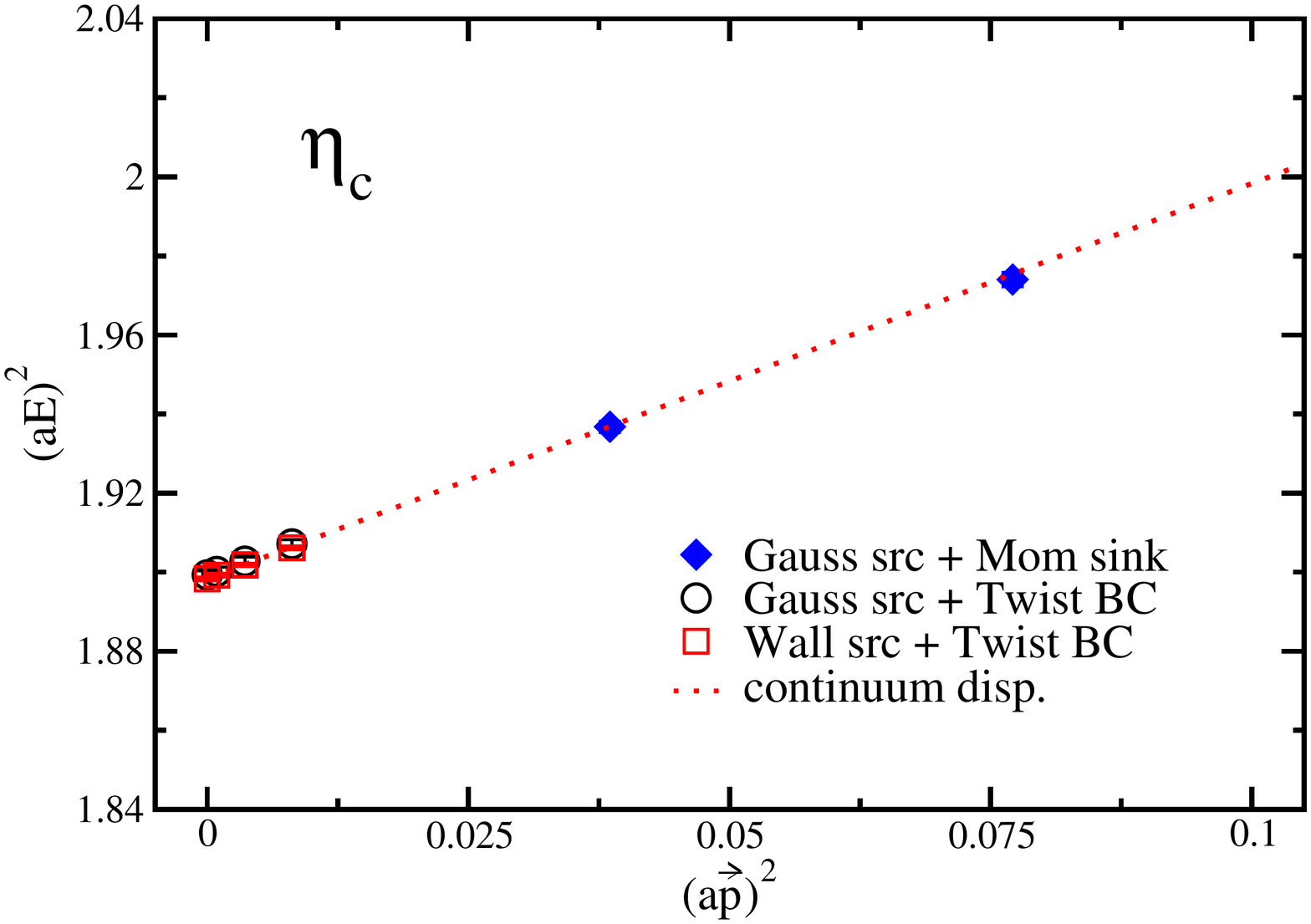}
  \includegraphics[width=.48\textwidth]{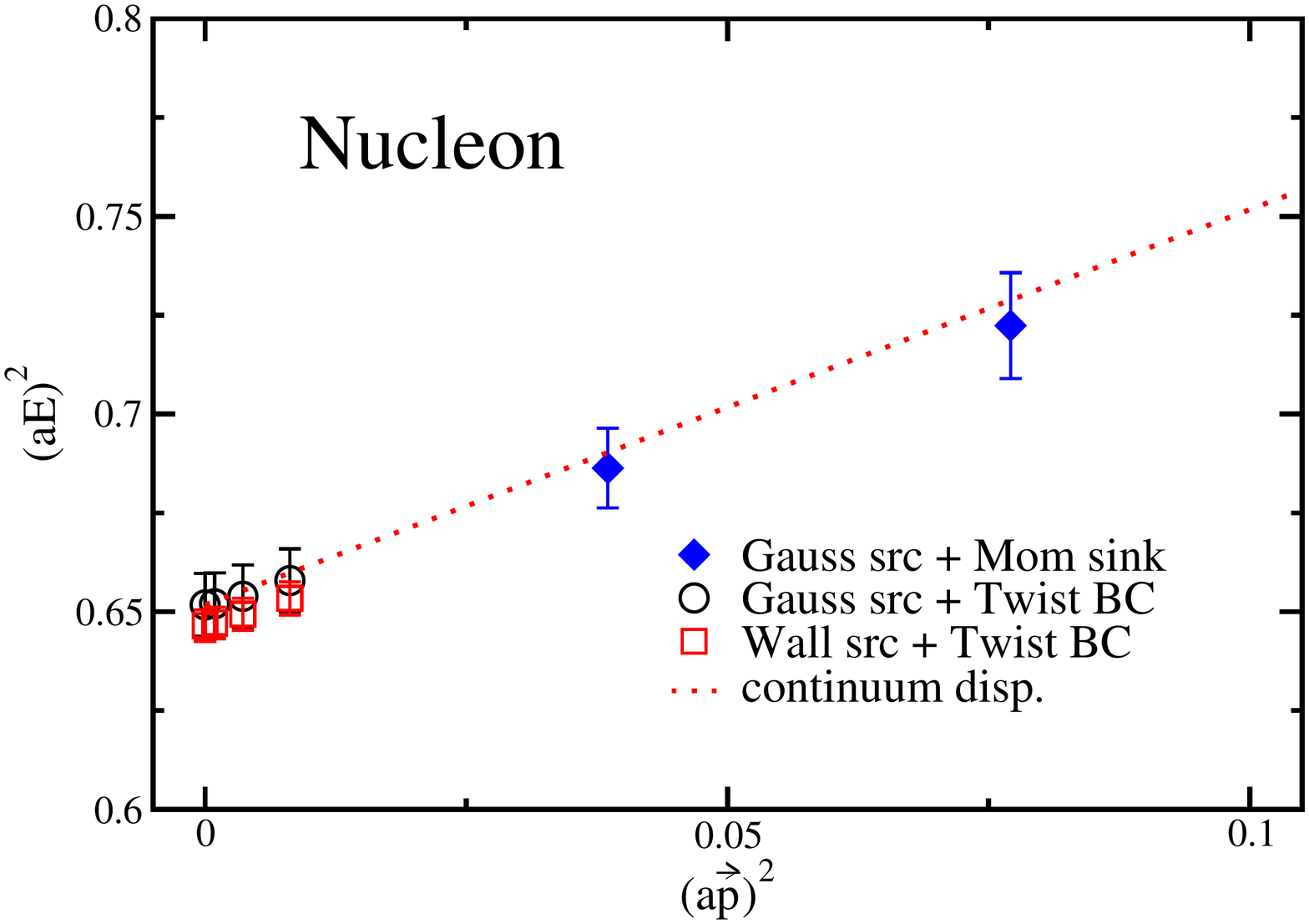} 
  \caption{The square of measured energies of the $\eta_c$ state (left) and the nucleon (right) 
  as a function of three momentum squared in lattice unit. 
  The dotted lines represents the relativistic and continuum-type dispersion relation.}
  \label{dispersion}
 \end{figure*}

As shown in Fig.~\ref{dispersion}, we observe that measured energies of the $\eta_c$ and the nucleon
under partially twisted boundary conditions increases as twisted angles increase. Here, 
the momentum squared $p^2$ can be evaluated by $(\varphi_3/L)^2$. 
Data points of energies calculated by the standard technique of the momentum projection 
using the Fourier transformation are also included for comparison. 
All data points are consistent with the relativistic and continuum-type dispersion relation $E_h^2=M_h^2+p^2$.
We recall that the dispersion relation is a key ingredient in determination of the relative momentum 
of two-particle states, which is required for L\"uscher's finite size method.

To determine the relative momentum of the charmonium-nucleon system, we consider
the interaction energy $E$, which is defined by an energy difference between the total energy
of two-hadrons and the sum of the rest mass of individual hadrons:
\begin{equation}
E=E_{c\bar{c}\hyn N}-(M_{c\bar{c}}+M_{N}).
\end{equation}
This energy value can be evaluated by the large-$t$ behavior of a ratio of 
the four-point correlation function $G_{c\bar{c}\hyn N}(t)=\sum_{\bf r}G^{c\bar{c}\hyn N}({\bf r}, t; t_{\rm src})$ and two-point correlation functions
of individual hadrons
\begin{equation}
   R_{c\bar{c}\hyn N}(t) = \frac{G_{c\bar{c}\hyn N}(t)}
    {G_{c\bar{c}}(t)G_{N}(t)}
    \ \ \ \xrightarrow[t \gg t_{\text{src}}] \ \ 
    \exp(-E(t-t_{\rm src}))
\end{equation}
where $G_{c\bar{c}}(t)$ and $G_{N}(t)$ represent two-point correlation functions of
the charmonium and nucleon, respectively.  
The interacting momentum $p$, which is defined in Eq.~(\ref{Eq:IntMom}), 
can be evaluated by this measured interaction energy $E$ with the rest masses of
individual hadrons, $M_{c\bar{c}}$ and $M_{N}$.
Therefore, if the four-point function $G_{c\bar{c}\hyn N}(t)$
is calculated under the twisted boundary conditions, we can get 
various interacting momenta $p$ near the threshold and also evaluate 
their corresponding scattering phase shifts through the extended L\"uscher's 
phase shift formula, which is described in Sec.~\ref{Sec:twist}.

  \begin{figure}
  \centering
   \includegraphics[width=.48\textwidth]{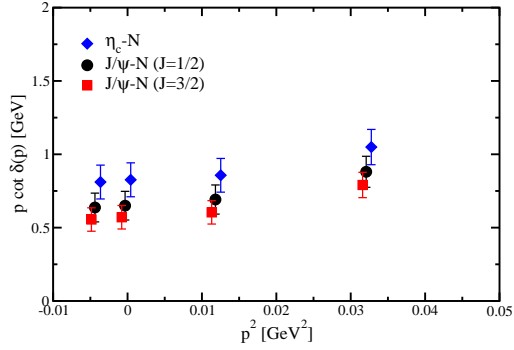} 
     \caption{The value of $p \cot \delta_0(p)$ as a function of the interaction momentum squared $p^2$
     at $M_{\pi}=0.64$ GeV.
     }
  \label{pcotD}
  \end{figure}

Fig.~\ref{pcotD} shows the value of $p\cot \delta_0(p)$ calculated at
the lightest quark mass ($M_{\pi}=0.64$ GeV) as a function of the interaction momentum squared $p^2$. 
Full diamond, circle and square symbols represent the $\eta_c\hyn N$, spin-1/2 $J/\psi\hyn N$ and
spin-3/2 $J/\psi\hyn N$ channels, respectively~\footnote{In the case of the $s$-wave $J/\psi\hyn N$
scattering, there are different spin states, spin-1/2 and 3/2 states.
Therefore, the appropriate spin projections are required to disentangle each spin
contribution from the four-point correlation functions. Details of the spin projection 
may be found in Appendix of Ref.~\cite{Yokokawa:2006td}.}.
All channels exhibit very mild momentum dependence near the threshold.
Analyticity of $p\cot \delta_0(p)$ in the vicinity of the threshold allows us to consider the 
fit ans\"atz as a simple polynomial function of the interacting momentum squared $p^2$:
\begin{equation}
p\cot \delta_0(p)= d_0 + d_1 p^2 + d_2 p^4.
\end{equation}
Needless to say, it is nothing but the effective-range expansion.
A linear fit with respect to $p^2$ for the three lowest $p^2$ points is 
enough to evaluate the first two fitting parameters, namely the scattering 
length $a_0=1/d_0$ and the effective range $r_0=2d_1$. We also apply a 
quadratic fit for all four data points.
The scattering parameters obtained from both determinations 
agree with each other within their errors. We simply choose the values obtained from 
the linear fit as our final results. 

\begin{figure*}
 \centering
 \includegraphics[width=.48\textwidth]{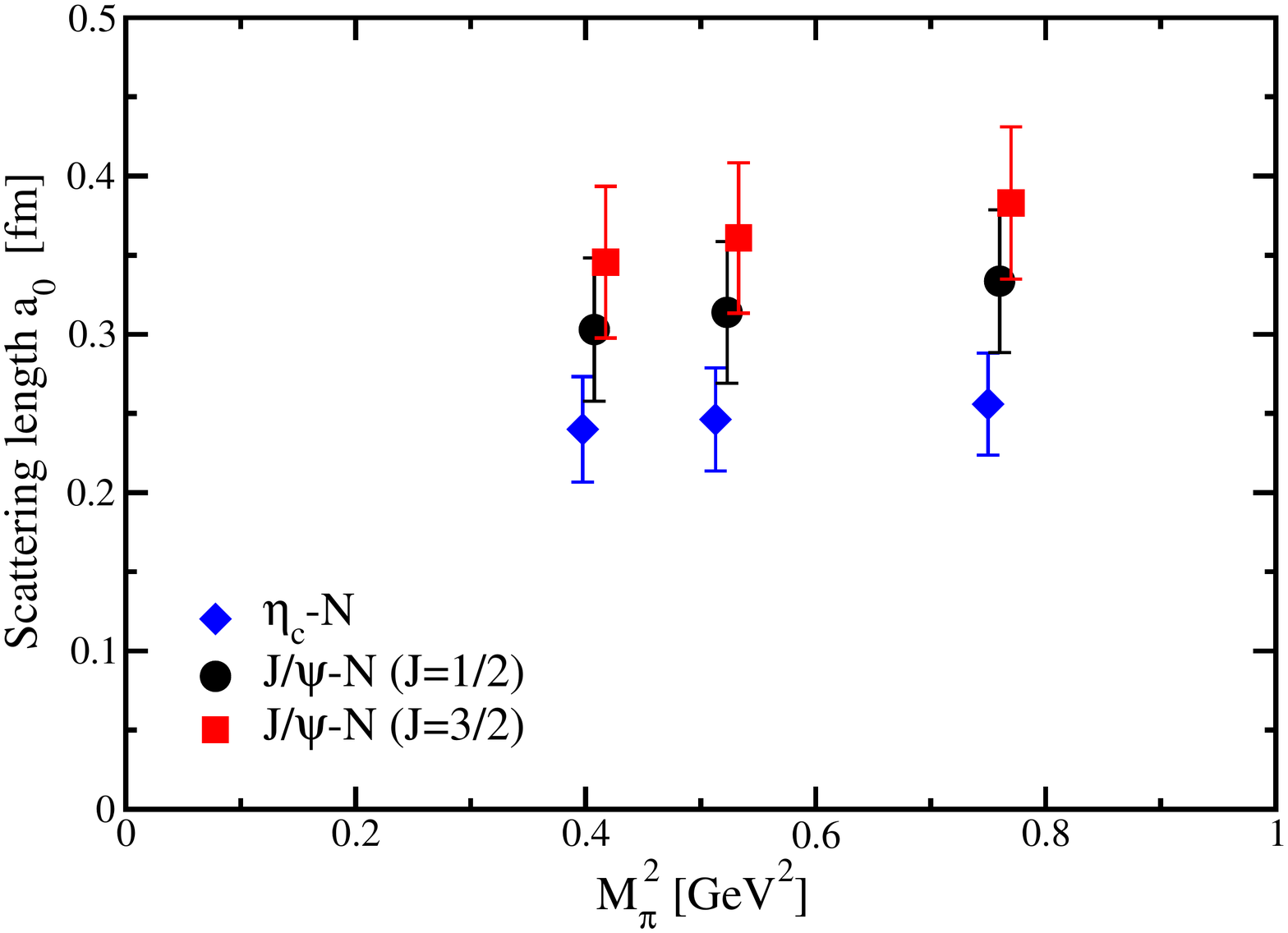} 
 \includegraphics[width=.48\textwidth]{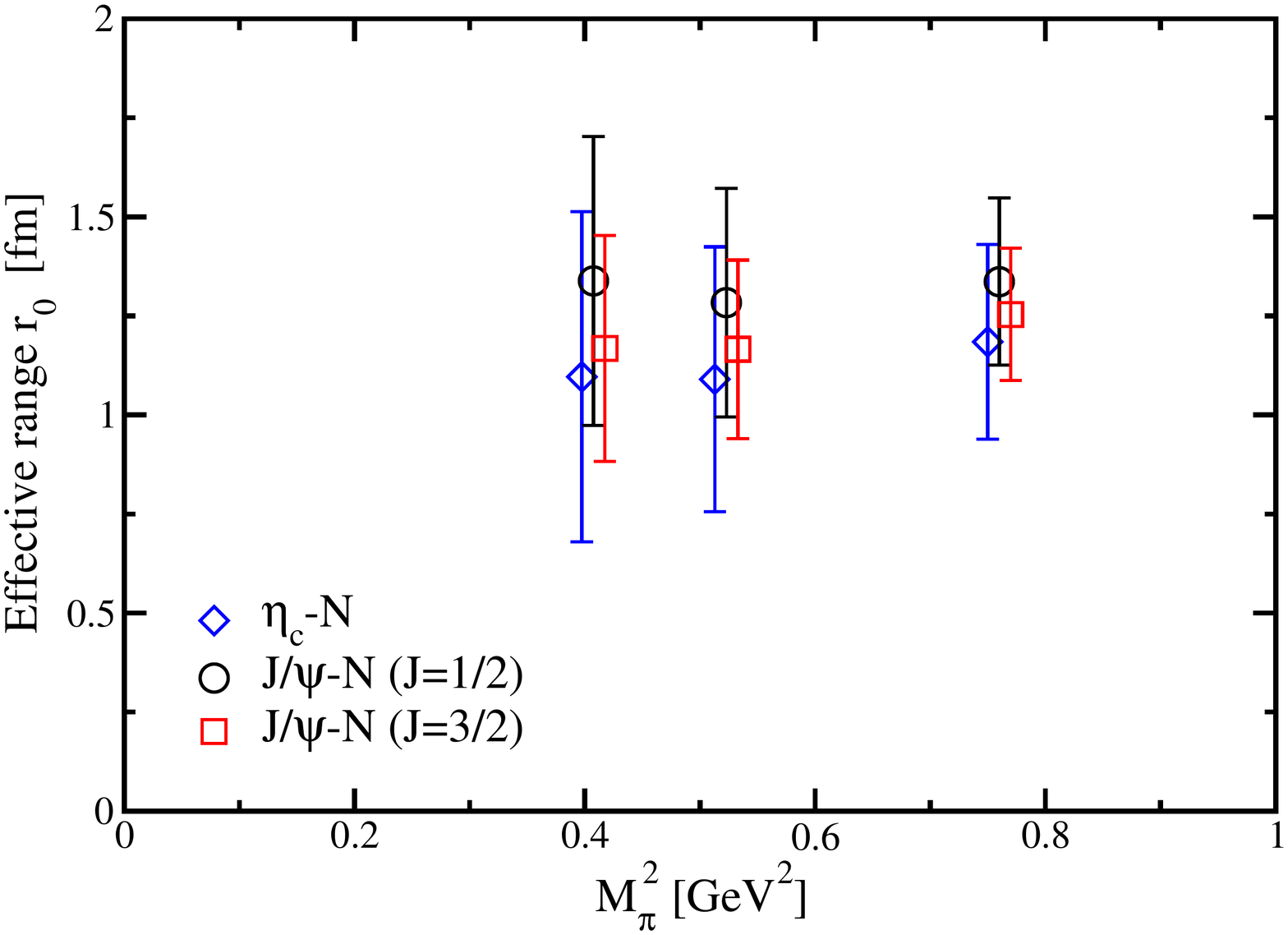}
 \caption{The scattering length $a_0$ (left) and effective range $r_0$ (right) as a function of $M_{\pi}^2$.  
 The squared (diamond) symbols have been moved slightly in the plus (minus)
 x-direction.
 }
 \label{scattering}
\end{figure*}

In Fig.~\ref{scattering}, we plot our evaluated scattering lengths (left panel) and effective ranges
(right panel) for all three channels as a function of pion mass squared. In both scattering parameters,
it is found that there is no significant quark mass dependence. 
These observations are consistent with what we observed in the charmonium-nucleon potentials. 
Although the channel dependence of the $c{\bar c}\hyn N$ scattering length was less clear 
in previous studies~\cite{Yokokawa:2006td,Liu:2008rza}, we find $(a_0^{J/\psi \hyn N})_{\rm SAV}\sim 0.35\;{\rm fm}  > a_0^{\eta_c \hyn N}\sim 0.25 \;{\rm fm}$~\footnote{
Here SAV stands for the spin-averaged value
$\frac{1}{3}[(a_{0})_{1/2}+2(a_{0})_{3/2}]$ for the $J/\psi\hyn N$ channel.} in this study.
On the other hand, both $\eta_c \hyn N$ and $J/\psi\hyn N$ channels yield the similar value of $r_0\sim  1.0\;{\rm fm}$ albeit with large errors.
The former feature may indicate that the $J/\psi\hyn N$ system is slightly more attractive than 
the $\eta_c\hyn N$ system at low energy. This is consistent with the similar spin dependence observed 
in the difference of the $\eta_c\hyn N$ and $J/\psi\hyn N$
potentials~\cite{Kawanai:2010ev}.

\section{Summary}
 We have studied low energy charmonium-nucleon interaction in both quenched
 and full QCD simulations. We first calculate potentials between the $\eta_c$ state 
 and the nucleon from the equal-time BS amplitude 
 through the effective Schr\"odinger equation. 
 We have found that the central potential
 $V_{c\bar{c}\hyn N}(r)$ in the $\eta_c$-$N$ system is weakly attractive at
 short distances and exponentially screened at large distances. 
 It is observed that the potential have no significantly large quark-mass dependence 
 within pion mass region $640\;\textrm{MeV}\leq M_\pi\leq 870\;\textrm{MeV}$ in quenched simulations.
 Our preliminary full QCD results show a good agreement with the quenched results. At least up to
 $M_{\pi}$=410 MeV, we observe no appreciable dynamical quark effect on the charmonium-nucleon potential.
 We have also employed an alternative approach for studying the charmonium-nucleon interaction.
 The $s$-wave phase shifts at low energies are calculated through the extended 
 L\"uscher's finite size method with twisted boundary conditions. We have successfully evaluated 
 both the scattering length and effective range from the charmonium-nucleon scattering phase shift
 in the vicinity of the threshold, where the effective range expansion is applicable.
 We have found $(a_0^{J/\psi \hyn N})_{\rm SAV}\sim 0.35\;{\rm fm}  > a_0^{\eta_c \hyn N}\sim 0.25 \;{\rm fm}$, 
 while all $\eta_c \hyn N$ and $J/\psi\hyn N$ channels yield the similar
 value of $r_0\sim 1.0\;{\rm fm}$ albeit with large errors. The channel dependence observed in 
 the $c{\bar c}\hyn N$ scattering length may indicate that the $J/\psi\hyn N$ system is slightly more 
 attractive than the $\eta_c\hyn N$ system at low energy. This is fairly consistent with 
 what we reported in Ref.~\cite{Kawanai:2010ev}, where the difference between 
 the $\eta_c\hyn N$ and $J/\psi\hyn N$ potentials are discussed.
 
 \section*{Acknowledgement}

 We would like to thank T. Hatsuda for helpful suggestions and fruitful discussions.
 We also thank PACS-CS Collaboration for their gauge configurations.
 T.K. is supported by Grant-in-Aid for JSPS Fellows (No.~22-7653).
 S.S. is  supported by the JSPS Grant-in-Aids for Scientific Research (C)
 (No.~19540265) and Scientific Research on Innovative Areas
 (No.~21105504). 
 Numerical calculations reported here were carried out
 on the PACS-CS supercomputer at CCS, University of Tsukuba and also on
 the T2K supercomputer at ITC, University of Tokyo.

\end{document}